\documentstyle[sprocl,epsfig]{article}

\def\Journal#1#2#3#4{{#1} {\bf #2}, #3 (#4)}

\def\PLB{{\em Phys. Lett.}  B}
\def\PRL{\em Phys. Rev. Lett.}
\def\PRD{{\em Phys. Rev.} D}

\def\be{\begin{equation}}
\def\ee{\end{equation}}
\def\bea{\begin{eqnarray}}
\def\eea{\end{eqnarray}}

\begin{document}

\title{HBT correlation in 158 A$\cdot$GeV Pb+Pb collisions}

\author{R. Ganz for the NA49 collaboration}

\address{Max-Planck-Institut f\"ur Physik M\"unchen, 
F\"ohringer Ring 6,\\ D-80805 M\"unchen, Germany\\E-mail: ganz@mppmu.mpg.de} 

\author{H.~Appelsh\"{a}user$^{7,\#}$, J.~B\"{a}chler$^{5}$,
S.J.~Bailey$^{17}$, D.~Barna$^{4}$, L.S.~Barnby$^{3}$,
J.~Bartke$^{6}$, R.A.~Barton$^{3}$, H.~Bia{\l}\-kowska$^{15}$,
A.~Billmeier$^{10}$, C.O.~Blyth$^{3}$
% ...
}

\address{
% ...
$^{3}$Birmingham University, Birmingham, England.\\
$^{4}$KFKI Research Institute for Particle and Nuclear Physics, Budapest, Hungary.\\
$^{5}$CERN, Geneva, Switzerland.\\
$^{6}$Institute of Nuclear Physics, Cracow, Poland.\\
$^{7}$Gesellschaft f\"{u}r Schwerionenforschung (GSI), Darmstadt, Germany.\\
% ...
$^{10}$Fachbereich Physik der Universit\"{a}t, Frankfurt, Germany.\\
% ...
$^{15}$Institute for Nuclear Studies, Warsaw, Poland.\\
% ...
$^{17}$Nuclear Physics Laboratory, University of Washington, Seattle, WA, USA.\\
% ...
$^{\#}$present address: Physikalisches Institut, Universitaet Heidelberg, Germany.\\
% ...
}

\maketitle\abstracts{ 
The large acceptance TPCs of the NA49 spectrometer allow
for a systematic multidimensional study of two-particle correlations 
in different part of phase space. Results from Bertsch-Pratt and 
Yano-Koonin-Podgoretskii parametrizations are presented differentially
in transverse pair momentum and pair rapidity.
These studies give an insight into the dynamical space-time evolution of 
relativistic Pb+Pb collisions, which is dominated by 
longitudinal expansion. }

\section{Introduction}

Recent high statistics experiments have demonstrated that, 
in heavy ion collisions at relativistic beam energies,
a single characteristic radius from intensity interferometry
does not exist.
First of all one deals with systems, in which the emission 
of particles is distributed over various ranges 
in three dimensional space as well as in time.
Secondly, due to the dynamical behaviour of the source 
(eg. collective expansion) 
correlations arise between the momenta of particles 
and the point of emission in space-time, which results
in a dependence of HBT radii on kinematical quantities.
In the case of a "boost invariant" picture, 
first suggested by Bjorken\cite{Bjorken}, 
the hot and dense initial stage of the collision expands longitudinally 
until freeze-out such, that a distinct longitudinal velocity ($v_z$)
profile is established: $v_z=z/t_f$. Here,
$z$ is the distance of particle emission at freeze-out
from the center of the collision and $t_f$ the freeze-out time. 
With respect to HBT correlations -- appearing at low momentum differences 
of particle pairs -- such a velocity profile disconnects different
parts of the source along the beam axis.
Only additional thermal motion with average velocities of 
$<v_{therm}>\approx\sqrt{T/m_\perp}$ can
-- over certain distances --  
compensate for the longitudinal velocity gradient. 
This means that, in case of relativistic heavy 
ion collisions, HBT radii do not measure the 
geometry of the source, but a ``length of homogeneity'' \cite{Sinyukov} 
$dz= R_z=t_f\sqrt{T/m_\perp}$, 
which may vary inside the source 
and hence depend on the phase space in which the
pairs are observed.

Experimentally, the question of space-time size of the source
is addressed by extracting HBT radii for all space-time 
components of the momentum difference vector $q^\mu =p^\mu_1 -p^\mu_2; \mu=0..3$
of pairs of identical particles 1 and 2.
Unfortunately, the ``on-mass-shell''- constraint for the observed (real) particles
reduces the degrees of freedom to three. 
For pairs this constraint can be written as
\begin{equation}\label{onmass}
q_0k_0 = \vec{q}\vec{k}
\end{equation}
with the average pair momentum $k^\mu=(p^\mu_1+p^\mu_2)/2$. 
Later it will be shown how this condition is utilized
to eliminate one of the four space-time components of $q^\mu$ to deduce 
parametrizations of the three dimensional correlation function $C_2(q)$ according to
Bertsch-Pratt(BP \cite{BP}) or to Yano-Koonin-Podgoretskii (YKP \cite{YKP}).

The second point -- the dynamical space momentum correlation --
is investigated experimentally by carrying out full three dimensional HBT analyses 
for intervals in the average transverse pair momentum ($k_\perp$) and average pair
rapidity ($y=(y_1+y_2)/2$), separately. 
All together this can be expressed by (q components introduced in section~\ref{Qsec}):
\begin{equation}
C_2(q^\mu)\longrightarrow C_2(q^\mu,k^\mu)=C_2\left(\left[{BP:q_{side},q_{out},q_{long} \atop 
                     YKP:q_{\perp},q_{0},q_{\|}}\right], \left[k_\perp,y\right]\right)
\end{equation}

Such a detailed analysis 
requires of course a large data sample
over a large fraction of phase space.
Facilitated by the large number of particles produced in central 
Pb+Pb
collisions at 158~A$\cdot$GeV 
and the large acceptance of the NA49 spectrometer at the CERN SPS 
accelerator such a study has become feasible and is presented here.  

\section{NA49 spectrometer and data analysis}

\begin{figure}
\begin{center}
\mbox{\epsfig{file=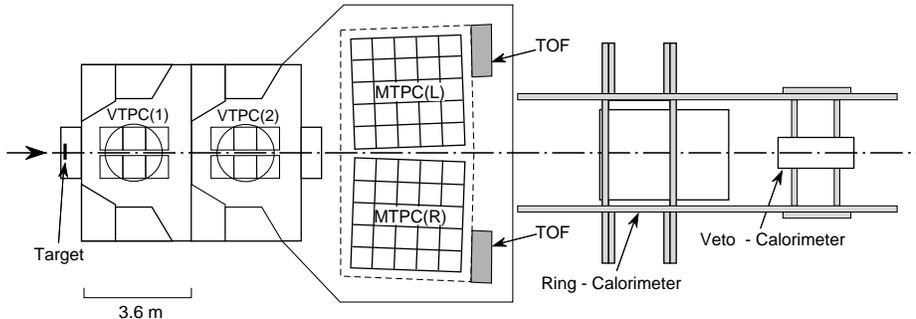,bbllx=-34pt,bblly=400pt,bburx=660pt,%
bbury=600pt,width=\linewidth}}               
\end{center}
\caption%
{\small The NA49 spectrometer}\label{NA49}
\end{figure}

The NA49 spectrometer\cite{Seyboth} (figure~\ref{NA49}) is located in the 
North area of the CERN SPS.
The results presented here are derived 
from the run period in 1995 with a 158~A$\cdot$GeV
Pb-beam. In every spill (5~s) about $10^5$ Pb$^{82+}$ 
ions impinge on a 224~mg/cm$^2$ Pb-target. 
1\% of these projectiles interact with the nuclei 
of the target foil. The most central 5\% (impact parameter $b=0-3.5$~fm)
of these collision are selected by the NA49 trigger utilizing a 
{\it Veto- Calorimeter}, which measures the energy carried
by the beam spectator fragments.
In such central collision about 1200 charged particles are produced, 
roughly 800 of them
are detected in at least one of the four 
{\it TPC}s (Time Projection Chambers) 
of NA49. 
At 2.0~m and 5.8~m downstream of the target two smaller
chambers ({\it VTPC1, VTPC2}; volume=$2.0\times 0.7\times 2.5$~m$^3$) are placed 
inside magnetic dipole fields.
In the ``Standard Configuration'' the fields
amount for 1.5~Telsa ({\it VTPC1}) and 1.1~Tesla ({\it VTPC2}), respectively, 
whereas in case of the ``Low field configuration'', 
fields of 0.3~Tesla and 1.5~Tesla have been used,
shifting the acceptance for pions towards central rapidities.
Further downstream -- outside the magnetic field --
the acceptance is extended to forward rapidities and higher momenta by two large volume TPCs' 
({\it MTPC}; volume=$3.8\times 1.2\times 3.8$~m$^3$). 

All four {\it TPC}s are equipped with a charge sensitive read-out, 
highly segmented along the beam (z-axis) and perpendicular to it 
in horizontal direction (x-axis).  
For each of these 180.000 pads
the charge is determined in 512 consecutive time slices
(100~ns), to determine the vertical (y-axis) position  
of particle tracks via the drift velocity ($v_d\approx$1.4~cm/$\mu$s inside {\it VTPC} 
and $v_d\approx$2.2~cm/$\mu$s inside {\it MTPC}).
This scheme allows for 3-dimensional track reconstruction.
From the curvature of reconstructed tracks inside the {\it VTPC}
the momentum of particles is determined
to a precision of $\delta p/p^2 \approx$~0.3\% in addition to their charge.
Moreover, all chambers measure the energy ($dE/dx$) deposited 
by the particle in the detector gas,
which -- in a future stage of analysis -- will be used to 
identify the particles. For the HBT analysis presented here, pairs of hadrons 
of the same charge ($h^-h^-$ or $h^+h^+$) without further identification
have been considered. 
These are dominated
by pairs of identical pions ($\pi^-\pi^-$ or $\pi^+\pi^+$). It has been shown, 
that a contamination from other particles has negligible influence on the determination of
HBT radii, with the exception of the chaoticy parameter $\lambda$, which will not be discussed
in this article.

Independent analyses have been carried out for the {\it VTPC2} and the {\it MTPC}s. 
The {\it VTPC2} analysis\cite{Appels} is based on 40.000 central Pb+Pb collisions, 
half of them taken in the ``Low field configuration'' and the other half in 
``Standard field configuration''. The analysis of the {\it MTPC} data\cite{Schoen} 
includes the same 40.000 events and adds another 50.000 events 
in ``Standard field configuration''. In both analyses pairs of tracks with 
distances less than 2~cm inside the TPC have been excluded to eliminate the 
influence of two particle reconstruction inefficiency 
for very close tracks. The same requirement has been imposed on
the distribution of uncorrelated pairs, which is
generated by combining tracks from different events. 
This distribution is used as reference (denominator) in the determination 
of the correlation function.
Both data sets have been corrected for Coulomb final state interaction based on
measured correlations of opposite-sign charged particles.
In case of the {\it VTPC2}--data this correction is carried out 
by the Gamov factor 
$G(q_{inv}); q_{inv}=\sqrt{-(p_1-p2)^2}$ 
modified by a dumping term 
to account for the finite size
of the source \cite{Kardia}:
\begin{equation}
C_{2,corr}^{--} = C_{2,meas.}^{--}\times 
\left((G(q_{inv})-1)e^{-q_{inv}/q_{eff}}+1\right)
\end{equation}
The parameter $q_{eff}$ is determined from a 
fit of the 1-dimensional correlation function 
$C_2^{+-}(q_{inv})$. 
 
The {\it MTPC}--data are corrected by the correlation function $C_2^{+-}$ 
of the opposite charged
particles, evaluated in the same three dimensionial relative momentum space 
as the like-sign pairs; eg. for BP-paramertization (see below): 
\begin{equation}
C_{2,corr}^{--} = C_{2,meas.}^{--}\times C^{+-}(q_{side},q_{out},q_{long})
\end{equation}

No further corrections have been applied to the data. The systematic error in the determination 
of the HBT radii is estimated to about 7\%.
The correlations function have been obtained
in different reference frames. The most intuitive choice 
of a reference frame for all pair momenta  
might be the {\bf C}enter- of {\bf M}ass {\bf S}ystem (CMS) 
of the colliding ions,
but in case of 
variations of the longitudinal velocity across the source
due to a longitudinal expansion, 
the {\bf L}ongitudinal {\bf C}o-{\bf M}oving {\bf S}ystem (LCMS) 
might be better suited.
It is defined on a pair-by-pair basis such that the longitudinal pair momentum $k_\|$
vanishes. 
The {\bf F}ixed {\bf L}ongitudinal {\bf C}o-{\bf M}oving {\bf S}ystem (FLCMS), 
in which the observer frame for every interval in pair rapidity 
is fixed at the center of that interval,
might be seen as a compromise between both.
In case of narrow widths of rapidity intervals LCMS and FLCMS are equivalent and are 
therefore treated as one in the following discussion, even though the 
{\it MTPC} data \cite{Schoen} have been evaluated in the CMS and FLCMS frames, whereas the
{\it VTPC2} data \cite{Appels} use the CMS and LCMS frames for BP projections and CMS and FLCMS
in case of YKP. 

\section{Q- parametrizations}\label{Qsec}
As shown in references \cite{Chapman,Heinz}, the correlation function for a source distribution,
expanded to second order at the
(in general $k$-dependent) space time points of maximum emission 
($\bar{X}$; saddle point) 
can be written as:
\begin{equation}\label{c2gauss}
C_2= 1+ e^{-q^\mu q^\nu <\hat{x}_\mu \hat{x}_\nu>}
\end{equation}
with $\hat{x}_\mu=x_\mu-\bar{X}_\mu$; $<>$ denotes an averaging over the source 
distribution. 
Even though this ``model independent'' expression (\ref{c2gauss}) 
can -- because of the constraint in eq. (\ref{onmass}) -- 
not be used to describe measured correlation functions, 
it is a generalization of both the Bertsch Pratt \cite{BP} as well as
the Yano Koonin Podgoretskii parametrization.
It therefore gains its importance as a tool when interpreting the radii
extracted in the framework of both formulations;
it furthermore provides a proof of consistency, when comparing results from both
parametriztions.
\begin{figure}
\begin{center}
\begin{minipage}{\linewidth}
\mbox{\epsfig{file=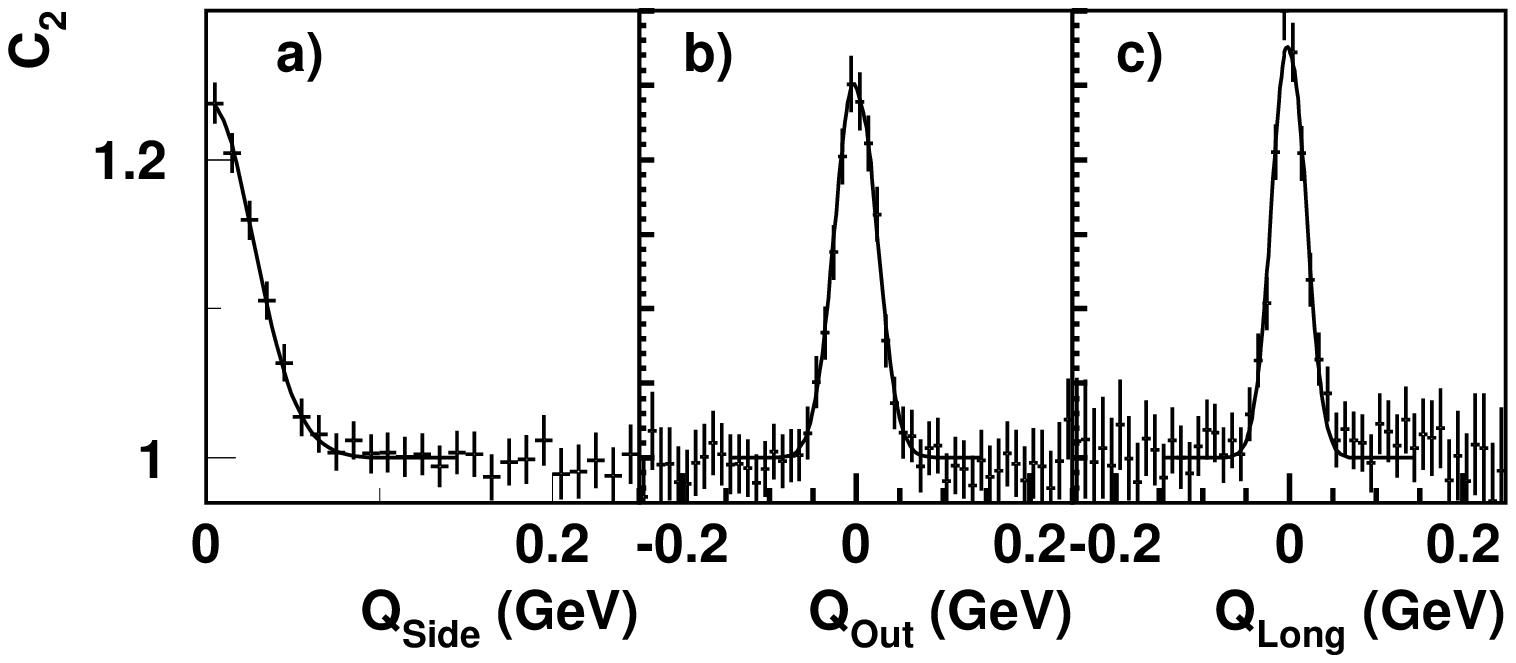,bbllx=0pt,bblly=0pt,bburx=500pt,%
bbury=190pt,width=0.8\linewidth}}
\end{minipage}

\begin{minipage}{\linewidth}
\mbox{\epsfig{file=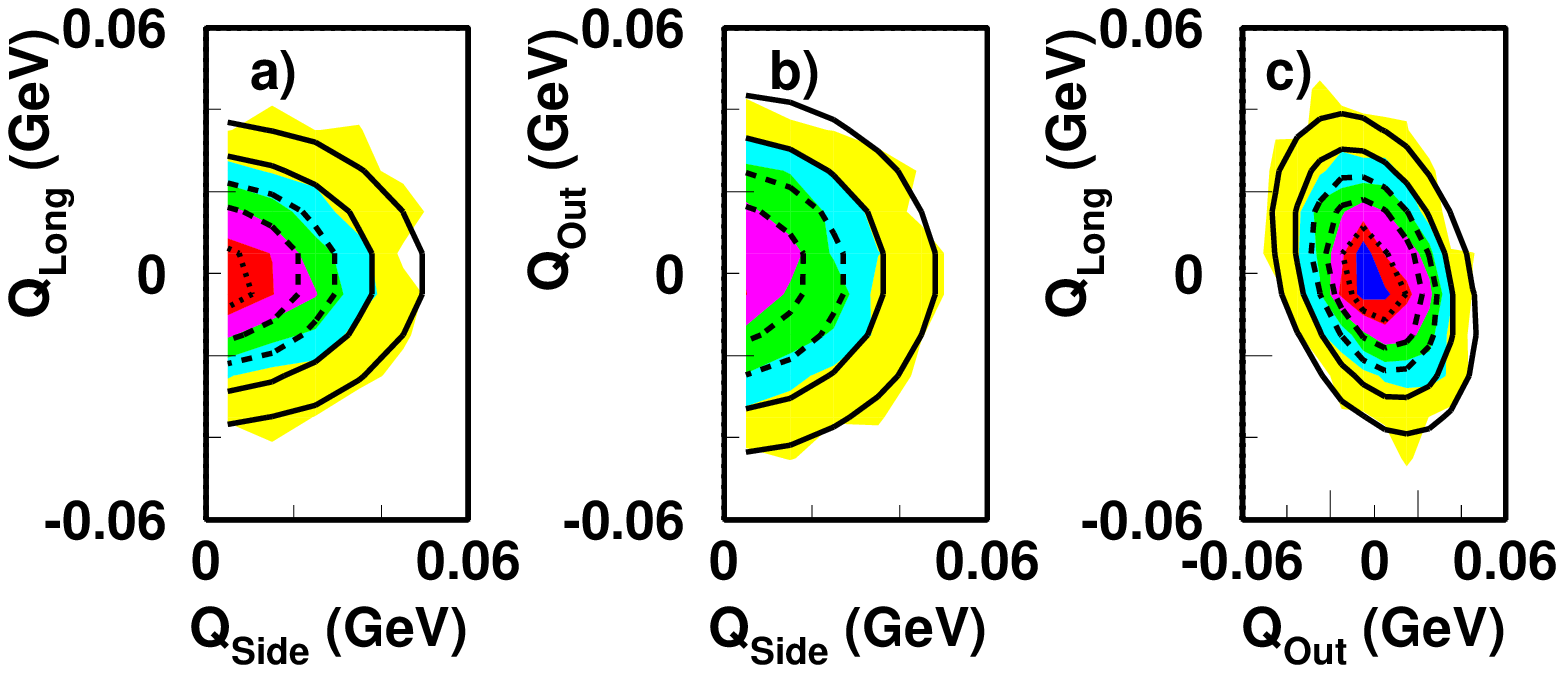,bbllx=0pt,bblly=60pt,bburx=500pt,%
bbury=190pt,width=0.8\linewidth}}
\end{minipage}
\end{center}
\caption%
{\small 1- and 2-dimensional projections of the $h^-h^-$
correlation function measured in the NA49 {\it MTPC}
using BP parameters for pair rapidities $4<y<5$ and transverse momenta 
$k_\perp<100$~MeV/c (FLCMS). 
The projections on components $q_j$ have been carried 
out by restricting the other components $q_i; i\ne j$ to $q_i<30$~MeV/c.}\label{BPproj}
\end{figure}
 
\subsection{BP parametrization}\label{BPsect}
Eliminating the temporal component of (\ref{c2gauss}) by condition (\ref{onmass}) in 
the form $q_0 = \vec{\beta}\vec{q}$ ($\vec{\beta}=\vec{k}/k_0$),
results in a correlation function parametrized according to reference~\cite{BP}:

\begin{equation}\label{BPfit}  
C_2 = 1+\lambda e^{-q^2_{side}R^2_{side}-q^2_{out}R^2_{out}-q^2_{long}R^2_{long} -
2q_{out}q_{long}R^2_{out-long}}
\end{equation}  
Here $q_{long}$ is the component of $q^\mu$ in beam direction, whereas
$q_{side}$ and $q_{out}$ are those perpendicular to it, with $q_{out}\|\vec{k}$  
and $q_{side}\perp\vec{k}$. 

\begin{figure}
\begin{center}
\mbox{\epsfig{file=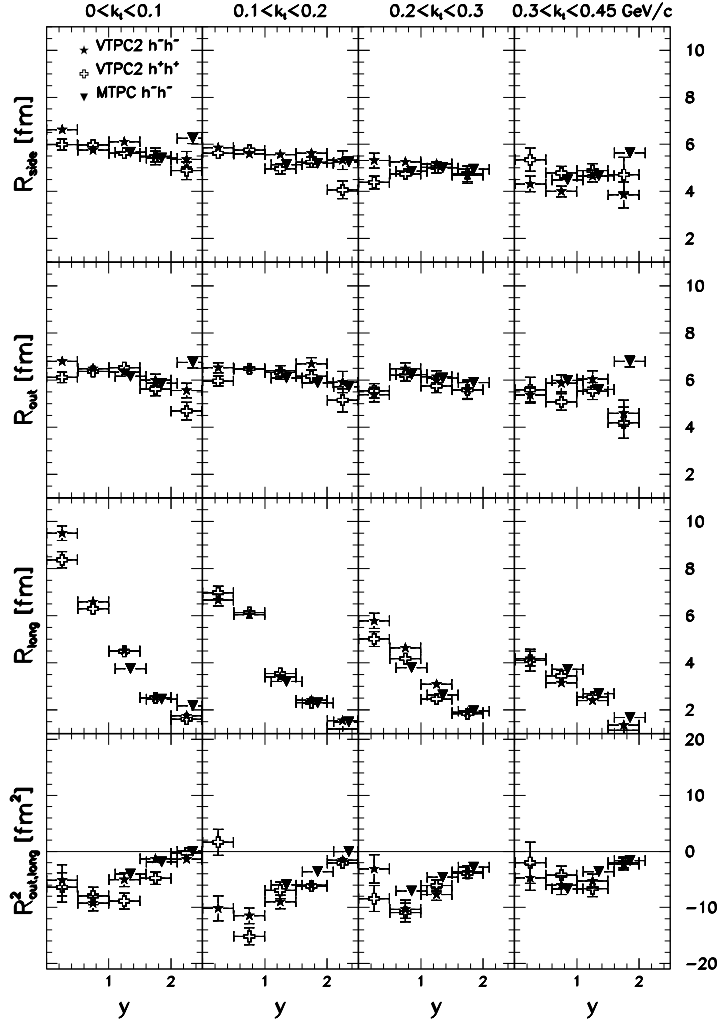,bbllx=0pt,bblly=30pt,bburx=650pt,%
bbury=700pt,width=\linewidth}}               
\end{center}
\caption%
{\small The dependence of BP radii on the pair rapidity for $h^-h^-$-pairs 
($\star$) and $h^+h^+$-pairs ({\bf $+$}) in the {\it VTPC2} and  
$h^-h^-$-pairs ($\bigtriangledown$) in the {\it MTPC}  
in four intervals of transverse momentum $k_t$ (CMS). The rapidity scale is shifted to the 
center of mass system of the ions.
The horizontal error bars correspond to the width of the intervals chosen in the analysis and the
vertical to the statistical errors only.
}\label{BPy}
\end{figure}

Due to symmetries of the sources considered here,  
only the ``out--long'' cross term remains, all others (``out--side'' and ``side--long'') vanish. 
This is consistent with the experimental observation presented in figure~\ref{BPy}
for $h^-h^-$-pairs at rapidities $4<y<5$ and transverse pair momentum $k_\perp<100$~MeV/c.
The contours lines in the 2-dimensional projection deviate from a circular shape (no cross term
correlation) only in case of the ``out--long'' projection. 
The  1-dimensional projections demonstrate the good agreement between the function
(\ref{BPfit}) and the data for small $q$, where the correlation signal is clearly visible, 
as well as for larger $q$, where the data points are consitent with $C_2=1$ within the 
error-bars.
At this point it should be emphasized, 
that all HBT radii presented here are derived by a simultaneous fit of all three components
of $q$ in $C_2$. The projections are generated for better visibility only.

When interpreting the BP radii it is advantageous to explicitly write
down the relation between equation~\ref{BPfit} and ansatz~\ref{c2gauss}:  
\begin{eqnarray}
R^2_{side}(\vec{k}) & = & <\hat{x}^2_y>\label{BPxyz1} \\
R^2_{out}(\vec{k}) & = & <(\hat{x}_x-\beta_\perp\hat{t})^2>\label{BPxyz2} \\ 
R^2_{long}(\vec{k}) & = & <(\hat{x}_z-\beta_\|\hat{t})^2> \\ 
R^2_{out,long}(\vec{k}) & = &<(\hat{x}_x-\beta_\perp\hat{t})(\hat{x}_z-\beta_\|\hat{t})> 
\end{eqnarray}
With the exception of $R_{side}$,
BP-radii mix spacial ($\hat{x}_{xyz}$)and temporal components ($\hat{t}$) of the source 
and 
an interpretation becomes therefore reference frame and model dependent.
In case of a longitudinally expanding source the ``length of homogeneity''
observed in the center of mass frame, appears Lorentz contracted 
in different intervals of pair rapidity. Such a behaviour is supported by the 
rapidity dependence of the measured 
radius $R_{long}$, as shown
for different intervals in $k_\perp$ in figure~\ref{BPy}. 
The data points can be described by $R_{long}=t_f/\mathrm{cosh}(y)\sqrt{T/m_\perp}$, 
which is the Lorentz frame dependent expression of section 1.
Assuming a temperature of $T=150$~MeV, a freeze-out time of $t_f\approx 9-7$~fm/c 
can be derived for the different $k_\perp$-intervals. 
By comparing 
the measured ``$side$''- and ``$out$''-radii to equation 
(\ref{BPxyz2})-(\ref{BPxyz1}) 
the duration time of freeze-out yields $\approx 2-4$~fm/c.
Both radii (``$side$'' and ``$out$'') appear to be constant over $y<2.5$ within errors.
In addition to the physical interpretation it is important to
note the good overall agreement between the different analyses 
and between pairs of opposite total charge, i.e. $h^-h^-$ compared to $h^+h^+$.


\subsection{YKP parametrization}

\begin{figure}
\begin{center}
\begin{minipage}{\linewidth}
\mbox{\epsfig{file=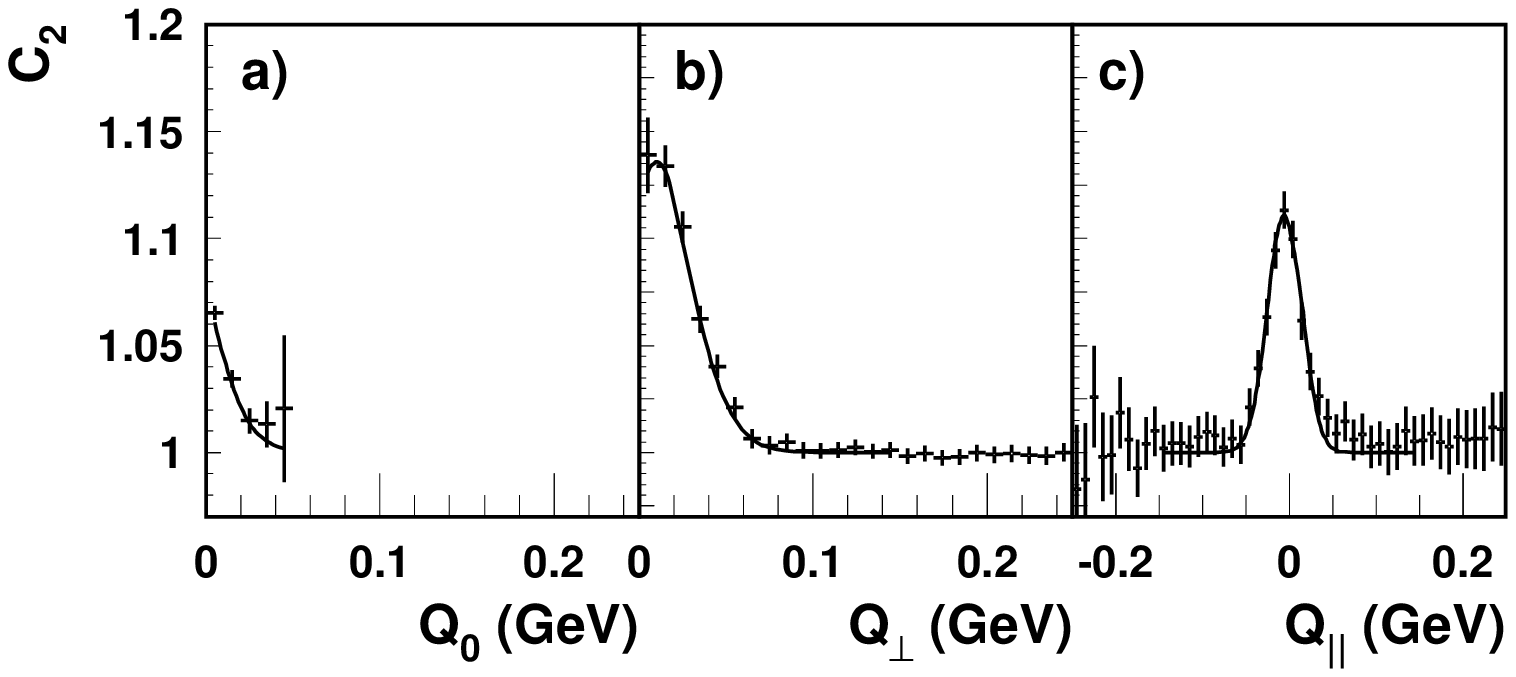,bbllx=0pt,bblly=0pt,bburx=500pt,%
bbury=190pt,width=0.8\linewidth}}
\end{minipage}

\begin{minipage}{\linewidth}
\mbox{\epsfig{file=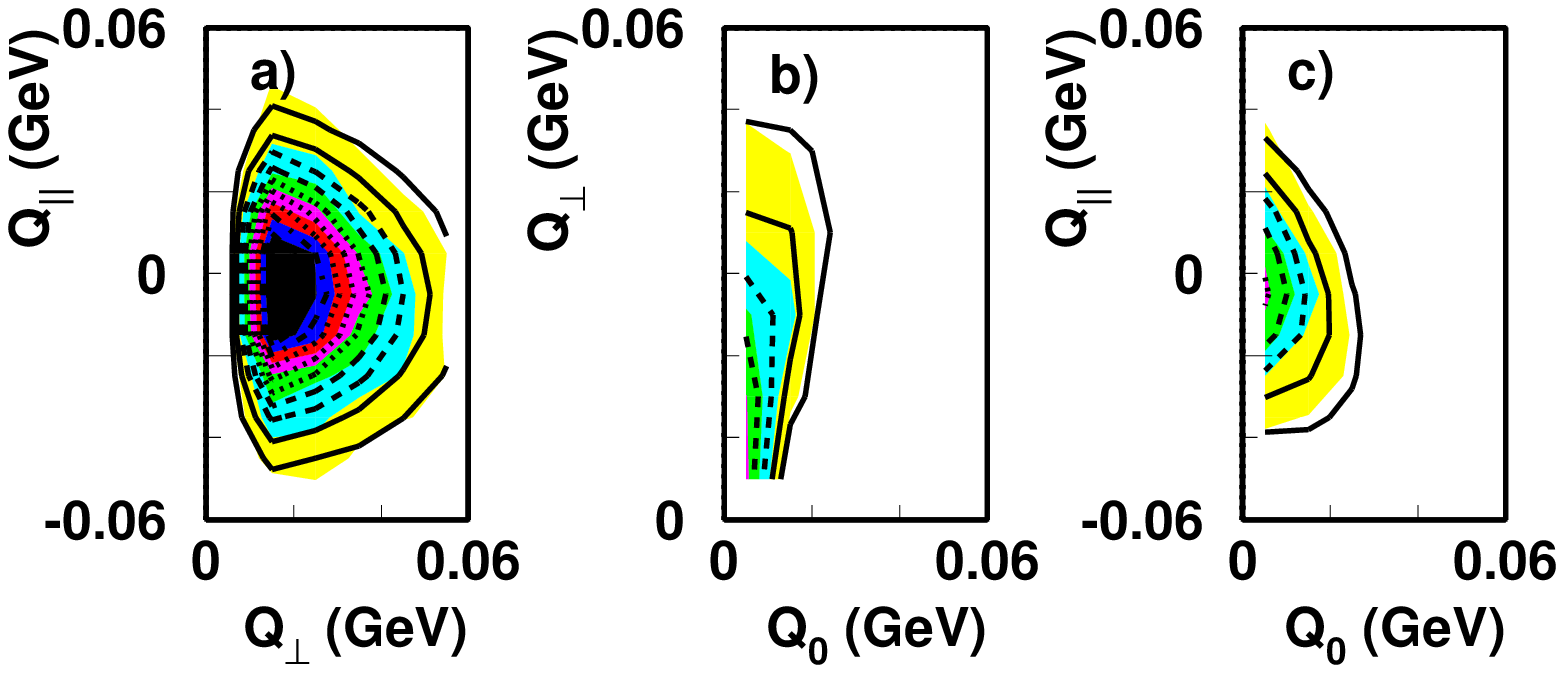,bbllx=0pt,bblly=50pt,bburx=500pt,%
bbury=190pt,width=0.8\linewidth}}
\end{minipage}
\end{center}
\caption%
{\small Projections of the same data sample as in figure~\ref{BPproj},
but this time in the YKP parametrization (FLCMS).
The projection onto component(s) $q_j$ have been carried 
out by restricting the other component(s) $q_i; i\ne j$ to $q_i<70$~MeV/c.  
}\label{YKPproj}
\end{figure}

Instead of eliminating the temporal component of (\ref{c2gauss}) in the BP formalism, 
condition (\ref{onmass}) might also be used  
via the relation $q_x =q_0/\beta_\perp - q_\|\beta_\|/\beta_\perp$.
With the picture of a boost invariant source in mind 
with different parts of the source moving at different longitudinal velocities, one
explicitly introduces a longitudinal velocity parameter ($\beta_{YKP}$) into the 
correlation function of YKP type \cite{YKP}:

\begin{equation}\label{YKPfit}  
C_2 = 1+\lambda e^{-q^2_{\perp}R^2_{\perp}
-\gamma^2_{YKP}(q_{\|}-\beta_{YKP}q_0)^2R^2_{\|}
-\gamma^2_{YKP}(q_{0}-\beta_{YKP}q_\|)^2R^2_{0}}
\end{equation}  
with $\gamma_{YKP}=1/\sqrt{1-\beta^2_{YKP}}$, ~~~$q_\perp=\sqrt{q^2_x+q^2_y}$ for the transverse momentum 
difference, $q_0$ for the energy difference and $q_\|$ for the longitudinal component.
In this case the interpretation of extracted radii becomes more evident, 
since space and time components are decoupled 
(the validity of the approximation is discussed in reference \cite{Wu}):  
\begin{eqnarray}\label{YKPxyz}
R^2_{\perp}(\vec{k}) & = & <\hat{x}^2_y>\label{YKPxyz1} \\
R^2_{0}(\vec{k}) & \approx & <\hat{t}^2>\label{YKPxyz2} \\ 
R^2_{\|}(\vec{k}) &\approx & <\hat{x}_z^2> 
\end{eqnarray}

\begin{figure}
\begin{center}
\mbox{\epsfig{file=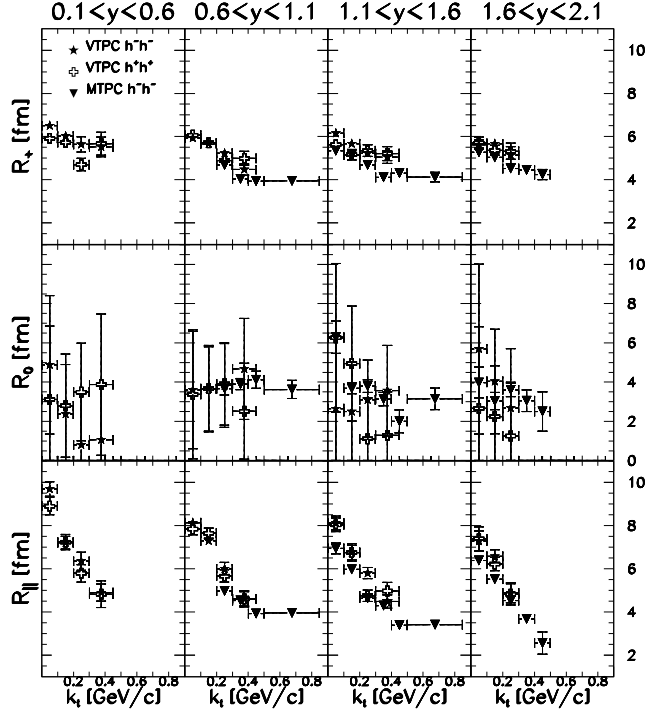,bbllx=34pt,bblly=230pt,bburx=530pt,%
bbury=760pt,width=0.7\linewidth}}               
\end{center}
\caption%
{\small The dependence of YKP-HBT radii on the transverse pair momentum
$k_\perp$ in different intervals of pair rapidity (LCMS) 
(otherwise same conventions as in figure~\ref{YKPkt}).
}
\label{YKPkt}
\end{figure}

Figure~\ref{YKPproj} shows one and 2-dimensional projections of the measured correlation function 
in YKP coordinates evaluated in the LCMS frame for the same $k_\perp$ and $y$-interval 
as in figure~\ref{BPproj}. Again, the data are described well by the chosen gaussian ansatz. 
A problem of this type of parametrization manifests itself in the projection of $q_0$. 
For a given interval in $q_\perp$ and $q_\|$ only a limited region of $q_0$ 
is kinematically available.
This is the reason for the large uncertainties of $R_0$ in figure~\ref{YKPkt}, 
which sumarizes the $k_t$-dependence for the YKP-radii in different intervals of rapidity. 
The clear decrease of $R_\|$ vs. $k_t$ again points to strong space-momentum correlations
in the source. Moreover, even in the transverse direction $R_\perp$ decreases 
for larger $k_t$-values,
which can be interpreted by transverse flow \cite{Appeletal}. 
The estimate of the duration of emission given in section~\ref{BPsect}
is confirmed by the extracted values of $R_0$. 

\begin{figure}
\begin{center}
\mbox{\epsfig{file=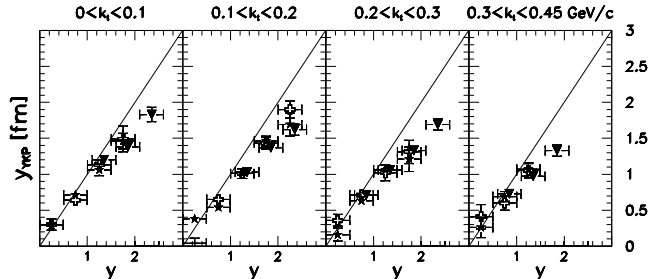,bbllx=34pt,bblly=595pt,bburx=530pt,%
bbury=760pt,width=0.7\linewidth}}               
\end{center}
\caption%
{\small 
Dependence of the Yano Koonin rapidity vs. the pair rapidity for four intervals in $k_\perp$ (LCMS).
}
\label{yYKP}
\end{figure}

In a parametrization of YKP type one can
gain further insight into the dynamics of the source
by utilizing the YKP velocity $\beta_{YKP}$. 
The YKP rapidity $y_{YKP}=\frac{1}{2}\ln\frac{1+\beta_{YKP}}{1-\beta_{YKP}}+y$
derived from the measured $\beta_{YKP}$
is compared to the pair rapidity in figure~\ref{yYKP}.
The data show the characteristics of a source, which expands 
in longitudinal direction. 
Even though deviations from an ideal boost invariant picture (line in figure~\ref{yYKP}), 
become apparant at forward rapidities, the consitstency which such a model is good. 

\section{Conclusion and Outlook}
Results from different analyses of multidimensional Bertsch-Pratt and Yano-Koonin-Podgoreskii
parametrizations of two particle correlation functions in 158~A$\cdot$GeV Pb+Pb
collisions have been presented differentially in pair rapidity $y$ and transverse pair momentum
$k_t$. 
The results confirm the importantance of extracting HBT-radii seperately
in different parts of phase space to disentangle the dynamical correlations of the source.
The pion source appears to expand in a close to boost invariant way, as seen in the rapidity
dependence of the Yano-Koonin velocity as well as of the $R_{long}$ radius in the CMS system.
Moreover, a finite duration time of emission of $2-4$~fm/c and a decreasing tranverse radius at large 
$k_t$ are observed. 
For a better understanding of the behaviour of the radii a comparison to a similar analysis 
of proton-proton and proton-lead collisions is currently in progress.  
Moreover, the centrality
dependence and beam energy dependence might constrain interpretations even further 
and will be investigated in upcoming analysis and further data taking.

\section*{Acknowledgments}
Acknowledgements: This work was supported by the Director, Office of Energy Research, 
Division of Nuclear Physics of the Office of High Energy and Nuclear Physics 
of the US Department of Energy under Contract DE-ACO3-76SFOOO98, 
the US National Science Foundation, 
the Bundesministerium fur Bildung und Forschung, Germany, 
the Alexander von Humboldt Foundation, 
the UK Engineering and Physical Sciences Research Council, 
the Polish State Committee for Scientific Research (2 P03B 01912), 
the Hungarian Scientific Research Foundation under contracts T14920 and T23790,
the EC Marie Curie Foundation,
and the Polish-German Foundation.


\section*{References}
\begin{thebibliography}{99}

\bibitem{Bjorken} J.D. Bjorken, \Journal{\PRD}{27}{140}{1983}.

\bibitem{Sinyukov} Yu. M. Sinyukov {\it et el.}, \Journal{{\em Nucl. Phys.} A}{498}{151}{1989}\\
and \Journal{{\em Nucl. Phys.} A}{566}{589}{1994}.


\bibitem{BP} G. F. Bertsch, \Journal{{\em Nucl. Phys.} A}{498}{173}{1989};\\
             S. Pratt, \Journal{{\em Phys. Rev.} D}{33}{1314}{1986}.  

\bibitem{YKP} F. B. Yano and S.E. Koonin, \Journal{\PLB}{78}{556}{1978};\\
         M.I. Podgoretskii, \Journal{{\em Sov. J. Nucl. Phys.}}{37}{272}{1983}\\
S. Chapman, P. Scotto and U. Heinz, \Journal{\PRL}{74}{4400}{1995}.

\bibitem{Seyboth} P. Seyboth {\it et al.}, 
{\em Proc. of XXV Int. Symposium on Multiparticle Dynamics} 
Stara Lesna Slovakia, 170 (1995).

\bibitem{Appels} H. Appelsh\"auser, PhD thesis, Univ. Frankfurt a.M. (1997).

\bibitem{Schoen} S. Sch\"onfelder, PhD thesis, MPI M\"unchen MPI-PhE/97-09 (1997).

\bibitem{Kardia} K. Kadija {\it et el.}, \Journal{{\em Nucl. Phys.} A}{610}{248}{1996}.

\bibitem{Chapman} S. Chapman {\it et al.}, \Journal{{\em Phys. Rev.} C}{52}{2694}{1995}.  

\bibitem{Heinz} U. Heinz {\it et al.}, \Journal{\PLB}{382}{181}{1996}.  
                
\bibitem{Wu} Y.F. Wu {\it et al.}, \Journal{{\em Euro. Phys. J.} C}{1}{599}{1998}.  

\bibitem{Appeletal} H. Appelsh\"auser {\it et al.}, 
\Journal{{\em Euro. Phys. J.} C}{2}{611}{1998}.  

\end{thebibliography}
\end{document}